\documentclass[aps,prl,reprint,amsmath,amssymb]{revtex4-2}
\usepackage{graphicx}
\usepackage{xcolor}
\usepackage{hyperref}
\usepackage{siunitx}
\usepackage{subcaption}
\usepackage{multirow}%
\usepackage{floatrow}
\usepackage[noend]{algpseudocode}
\usepackage[font=small,skip=3pt]{caption}
\usepackage{makecell}
\usepackage{textcomp}%
\usepackage{cleveref}
\usepackage{url}
\usepackage{tikz}
\usepackage{booktabs}
\usepackage{pgfplots}
\pgfplotsset{compat=1.18}
\usepackage[version=4]{mhchem}
\let\mrm\mathrm

\begin{document}

\title{The Value and Cost of Fusion Neutrons}
\author{J. F. Parisi}
\email{jason@marathonfusion.com}
\author{K. Schiller}
\affiliation{Marathon Fusion, 150 Mississippi, San Francisco, CA 94107, USA}

\begin{abstract}
Deuterium-tritium fusion reactions produce high-energy neutrons that can transmute materials into valuable isotopes. Over the next ten years, the cost of fusion neutrons is projected to decrease by roughly seven orders of magnitude. Most ($\sim$5 orders of magnitude) is technological overhang driven by the low availability of current experiments; the remaining $\sim$2 orders of magnitude require higher plasma gain and lower capital intensity. We introduce the levelized cost of a neutron (LCON), an economic metric analogous to the levelized cost of energy that gives the minimum neutron value for economic breakeven of a fusion system. LCON depends on plasma gain, capital intensity, availability, and neutron flux, and is offset by revenue from co-produced electricity, precious metals, and radioisotopes. The revenue per neutron spans at least ten orders of magnitude, from electricity and gold ($\sim$\$$10^{-20}$/neutron) to actinium-225 ($\sim$\$$10^{-10}$/neutron), defining a `neutron ladder': a staged, revenue-positive development pathway from current fusion devices to terawatt-scale power plants. 
\end{abstract}

\maketitle

From the earliest lamps to LEDs, the real cost of artificial light has fallen by roughly five orders of magnitude~\cite{nordhaus1996real,smil2018energy,potter2025origins}. Much of this came from a single paradigm shift - fires, candles, and oil lamps to electric incandescent bulbs - while the remaining factor of $\sim$100$\times$ came from sustained improvement within the new technology, from carbon filaments ($\sim$1-4\,lumen/W) to modern LEDs ($\gtrsim$100\,lumen/W). High-energy fusion neutrons are at the beginning of an analogous trajectory, albeit on a much accelerated timeline. Today's neutron sources produce 14.1\,MeV neutrons at $\gtrsim$\$\,$10^{-13}$ each; commercial-scale fusion promises to reduce this by seven orders of magnitude. Most of this gap is technological overhang - the low availability and high capital intensity of current experiments - rather than a fundamental limitation on fusion plasma performance. And just as candlelight was valuable long before it was `cheap,' fusion neutrons can generate significant socioeconomic value at every stage of cost reduction, from medical isotopes today to gold and electricity at scale. This Letter quantifies this cost trajectory and the value it unlocks.

Deuterium-tritium (D-T) fusion reactions produce high-energy \qty{14.1}{MeV} neutrons at a rate
\begin{equation}
    \dot{N}_\mathrm{n} = \frac{P_\mrm{fus}}{E_\mrm{fus}} \;\;\;\;\; [\mathrm{s}^{-1}],
    \label{eq:Ndot}
\end{equation}
where $P_\mrm{fus}$ is the fusion power and $E_\mrm{fus} = 2.8 \cdot 10^{-12}$\,J is the total energy per reaction~\cite{Wurzel2022}. Conventionally, these neutrons breed tritium and deposit energy in a blanket for conversion to electricity, contributing \$$\sim10^{-20}$ per neutron at current electricity prices. But fusion neutrons can also transmute feedstock materials into isotopes whose value per neutron exceeds the electricity value by many orders of magnitude~\cite{engholm1986radioisotope,rutkowski2025scalable,parisi2025isotope,parisi2025j,evitts2025theoretical,parisi2025isotopemuon}. This raises a basic economic question: what does a fusion neutron cost, how does its cost scale, and what are the limits on its cost?

We introduce the levelized cost of a neutron (LCON), analogous to the levelized cost of energy (LCOE) used in the electricity sector \cite{branker2011review,kost2013levelized,obi2017calculation,hansen2019decision,sens2022capital,schwartz2023value}. Just as the LCOE gives the minimum electricity price for a power plant to break even, the LCON gives the minimum revenue generated per neutron for a fusion system to achieve economic breakeven. We show that current neutron sources operate at costs at least seven orders of magnitude above commercial fusion targets, identify the factors that close this gap, and describe how the range of neutron values defines a `neutron ladder' providing a revenue-positive development pathway for fusion energy. While fusion neutrons are currently expensive to produce, they have valuable applications that subsidize neutron production and allow a high LCON to be tolerated.

\textit{Value per neutron.} In this work we consider two non-exclusive ways of generating value with fusion neutrons: electricity and isotopes. Each neutron deposits thermal energy $\mathcal{K} E_\mrm{fus}$ in the blanket, where $\mathcal{K} \approx 1.1$ is the blanket thermal multiplication~\cite{sawan_physics_2006} and we have included the alpha particle energy in $E_\mrm{fus}$. The value per neutron of sold electricity is
\begin{equation}
    v_\mathrm{n}^\mathrm{elec} = \eta \, \mathcal{K} \, E_\mrm{fus}\, \tilde{C}_e \;\; [\$],
    \label{eq:vn_elec}
\end{equation}
where electricity conversion efficiency is $\eta \approx 0.35$ and electricity price is $\tilde{C}_e$ (\$/J). For $C_e =$\$100/MWh$_e$, $v_\mathrm{n}^\mathrm{elec} \approx \$3.0\cdot10^{-20}$/n. In the long term, the cost of electricity will need to decrease substantially for fusion electricity to be competitive; we use \$100/MWh$_e$ as a representative near-term industrial electricity price and note its sensitivity in Table~\ref{tab:vn}. We have not subtracted the recirculating facility power in the neutron revenue in \cref{eq:vn_elec}, which would reduce the neutron value.

\begin{table}[tb]
\centering
\caption{Value per neutron $v_\mathrm{n}$ (\cref{eq:vn,eq:vn_elec}) for representative products. Isotope prices $C_\mrm{pro}$ and transmutation efficiencies $\eta_\mrm{pro}$ from Ref.~\cite{parisi2025isotope}. `\ce{^{197}Au} + Electricity' has an electricity price of $\$50$/MWh$_e$.}
\label{tab:vn}
\begin{ruledtabular}
\begin{tabular}{lccc}
Product & $C_\mrm{pro}$ (\$/kg) & $\eta_\mrm{pro}$ & $v_\mathrm{n}$ (\$/n) \\
\hline
Electricity ($\$50$/MWh$_e$) & & & $1.5\cdot10^{-20}$ \\
Electricity ($\$100$/MWh$_e$) & & & $3.0\cdot10^{-20}$ \\
\ce{^{197}Au} & $1.6\cdot10^{5}$ & 0.50 & $2.6\cdot10^{-20}$ \\
\ce{^{197}Au} + Electricity & $1.6\cdot10^{5}$ & 0.50 & $4.1\cdot10^{-20}$ \\
\ce{^{147}Pm} & $1.0\cdot10^{6}$ & 0.50 & $1.2\cdot10^{-19}$ \\
\ce{^{99}Mo} & $1.0\cdot10^{11}$ & $1.4\cdot10^{-3}$ & $2.6\cdot10^{-17}$ \\
\ce{^{225}Ac} & $5.0\cdot10^{14}$ & 0.50 & $9.3\cdot10^{-11}$ \\
\end{tabular}
\end{ruledtabular}
\end{table}

If a fraction $\eta_\mrm{pro}$ of fusion neutrons transmute a feedstock nucleus into a product of mass $m_\mrm{pro}$ and price $C_\mrm{pro}$ (\$/kg), the revenue generated per neutron is~\cite{parisi2025isotope}
\begin{equation}
    v_\mathrm{n}^\mrm{prod} = \eta_\mrm{pro}\, m_\mrm{pro}\, C_\mrm{pro}.
    \label{eq:vn}
\end{equation}
\Cref{tab:vn} lists $v_\mathrm{n}$ for representative products. The range spans ten orders of magnitude, from gold and electricity at $\sim\!10^{-20}$\,\$/n to \ce{^{225}Ac} at $\sim\!10^{-10}$\,\$/n. The same 14.1\,MeV neutron can be worth vastly different amounts. 

Transmutation products need not come at the expense of tritium self-sufficiency \cite{Abdou2021,Meschini2023}. Many pathways, such as $({\rm n},2{\rm n})$ reactions on mercury to produce gold, are themselves neutron-multiplying and compatible with tritium breeding~\cite{rutkowski2025scalable}. Others with lower cross sections, such as $({\rm n},\alpha)$ and $({\rm n},{\rm p})$ reactions for medical isotope production, are barely parasitic on the tritium breeding ratio and may coexist with a breeding blanket~\cite{parisi2025isotope}, although smaller MW-scale devices externally fueled with tritium may also serve as dedicated isotope production facilities.

\textit{Levelized cost of a neutron.} We define the LCON as the minimum neutron value required for a fusion transmutation plant to break even, that is, to achieve zero net present value (NPV) over its operating lifetime. A plant is profitable when revenue per neutron exceeds LCON.

\begin{figure}[bt!]
    \centering
    \begin{subfigure}[t]{\textwidth}
    \centering
    \includegraphics[width=0.97\textwidth]{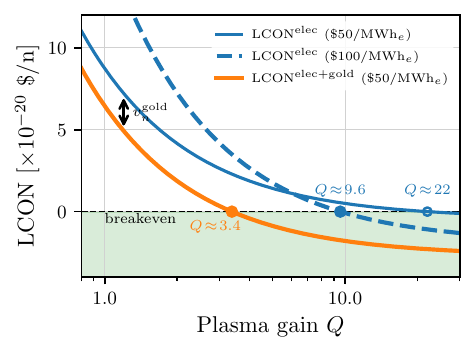}
    \caption{}
    \end{subfigure}
    \centering
    \begin{subfigure}[t]{\textwidth}
    \centering
    \includegraphics[width=0.97\textwidth]{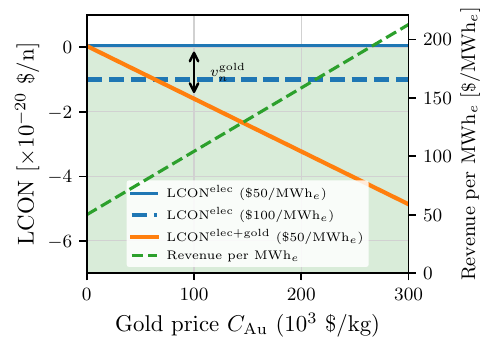}
    \caption{}
    \end{subfigure}
    \caption{(a) Effective LCON versus $Q$ at $I^\mathrm{cap} = \$2$\,B/GW and $\mathcal{A} = 1$. Blue curves: $\mathrm{LCON}^\mathrm{elec}$ at $C_e = \$50$/MWh$_e$ (solid) and $\$100$/MWh$_e$ (dashed), with electricity-only breakeven at $Q \approx 22$ and $\approx 10$, respectively. Orange: $\mathrm{LCON}^{\mathrm{elec+gold}}$ at $\$50$/MWh$_e$, with gold co-generation reducing breakeven to $Q \approx 3$. Green shading: profitable region. (b) LCON and revenue per MWh$_e$ (at \$50/MWh$_e$) versus gold price at $Q = 20$; blue lines show $\mathrm{LCON}^\mathrm{elec}$ at both electricity prices.}
\label{fig:LCON_cogen}
\end{figure}

Consider a fusion plant with capital cost per power $I^\mrm{cap}$ (\$/GW of thermal fusion power), plasma gain $Q=P_\mathrm{fus}/P_\mathrm{heat}$ where $P_\mathrm{heat}$ is the external plasma heating, availability factor $\mathcal{A}$ (fraction of the year the plant operates), and operating lifetime $L$ at discount rate $r$. The plant generates electricity from heat and pays for recirculating power $P_\mrm{circ}$ to sustain fusion reactions and operate systems. Setting $\mrm{NPV} = 0$ and solving for the breakeven neutron value (details in Appendix) gives~\cite{parisi2025isotope}
\begin{equation}
    \mrm{LCON} = c_\mathrm{n}^\mrm{cap} + c_\mathrm{n}^\mrm{op} \;\; \text{[\$/n]},
    \label{eq:LCON}
\end{equation}
with a capital component
\begin{equation}
    c_\mathrm{n}^\mrm{cap} = \frac{I^\mrm{cap}\, E_\mrm{fus}}{10^9\, T_\mrm{year}\, \mathcal{A}\, S_\mrm{disc}},
    \label{eq:cn_cap}
\end{equation}
where $T_\mrm{year} = 3.16 \cdot 10^7$\,s and $S_\mrm{disc} = \sum_{t=0}^{L}(1+r)^{-t}$ is the present-value discount sum, and an operating component representing the electricity cost of making a neutron,
\begin{equation}
    c_\mathrm{n}^\mrm{op} = \frac{E_\mrm{fus}\, \tilde{C}_e}{\eta_\mrm{heat} Q}  \left(\frac{1}{\eta_\mrm{abs} \, f_h} - \eta \right).
    \label{eq:cn_op}
\end{equation}
The first term is the electricity cost of sustaining fusion reactions, and the second is the recovery of heating-system waste heat through a thermodynamic cycle. The widely used metric `engineering breakeven'~\cite{Wurzel2022} occurs when $c^\mathrm{op}_\mathrm{n} = v_\mathrm{n}^\mathrm{elec}$. We neglect non-electricity operating costs in \Cref{eq:cn_op} for simplicity and because reliable data for fusion systems are lacking.

The capital component is independent of plasma performance but inversely proportional to availability: for $I^\mrm{cap} = \$2$\,B/GW with $\mathcal{A} = 1$, $L = 30$\,yr, and $r = 0.05$, we find $c_\mathrm{n}^\mrm{cap} \approx \$1.1\cdot10^{-20}$/n, about a third of $v^\mathrm{elec}_\mathrm{n}$; at $\mathcal{A} = 0.5$ the capital cost per neutron doubles. The operating component is dominated at low gain by the $1/Q$ term (recirculating power cost) and becomes negative at high gain as the plant exports electricity. The electricity revenue credit $v_\mathrm{n}^\mathrm{elec}$ offsets LCON, giving an effective LCON,
\begin{equation}
\mrm{LCON}^\mathrm{elec} = \mathrm{LCON} - v_\mathrm{n}^\mathrm{elec}.
\end{equation}
Substituting values for standard parameters~\cite{parisi2025isotope} ($\eta_\mrm{abs} = 0.95$, $\eta_\mrm{heat} = 0.60$, $f_h = 0.60$, $\eta = 0.35$, $\mathcal{K} = 1.1$, $C_e = \$100$/MWh$_e$) in $c_\mathrm{n}^\mrm{op}$ (\cref{eq:cn_op}), $\mrm{LCON}^\mathrm{elec}$ simplifies to
\begin{equation}
    \mrm{LCON}^\mathrm{elec} \approx \frac{1.8\cdot10^{-19}}{Q} + \frac{c_\mathrm{n}^{\mrm{cap},0}}{\mathcal{A}} - v_\mathrm{n}^\mathrm{elec},
    \label{eq:LCON_simple}
\end{equation}
where $c_\mathrm{n}^{\mrm{cap},0} \equiv I^\mrm{cap} E_\mrm{fus}/(10^9\, T_\mrm{year}\, S_\mrm{disc})$ is the capital cost per neutron at $\mathcal{A} = 1$. Reading left to right in \cref{eq:LCON_simple}: the first term is the recirculating power cost ($\propto 1/Q$), the second is amortized capital ($\propto 1/\mathcal{A}$), and the third is the revenue the plant earns by selling blanket heat as electricity. Crucially, $\mrm{LCON}^\mathrm{elec}$ must be negative for an electricity-only fusion system to be economically viable.

A fusion plant that co-produces transmutation products and electricity offsets LCON through the combined revenue of all products. The effective LCON including all revenue streams is
\begin{equation}
    \mrm{LCON}^{\mrm{elec+prod}} = \mrm{LCON}^{\mrm{elec}} - \sum_i v_\mathrm{n}^\mathrm{prod,i},
    \label{eq:LCON_general}
\end{equation}
where the sum runs over all transmutation products per source neutron. Each product revenue acts as an additional subsidy on the cost of neutrons. For example, co-producing electricity and gold gives
\begin{equation}
\mrm{LCON}^{\mrm{elec+gold}} = \mrm{LCON}^{\mrm{elec}} - v_\mathrm{n}^\mathrm{gold}.
\label{eq:LCON_Au}
\end{equation} 
\Cref{fig:LCON_cogen}(a) compares $\mrm{LCON}^\mrm{elec}$ at two electricity prices and $\mrm{LCON}^{\mrm{elec+gold}}$ at $C_e = \$50$/MWh$_e$, all at $I^\mrm{cap} = \$2$\,B/GW. Electricity-only breakeven occurs at $Q \approx 22$ (\$50/MWh$_e$) and $Q \approx 10$ (\$100/MWh$_e$); gold co-production at \$50/MWh$_e$ reduces breakeven further to $Q \approx 3$. Below this gain, a co-generating plant's electricity and gold revenue together exceed all costs, and any additional product revenue is profit. \Cref{fig:LCON_cogen}(b) shows how LCON and the plant revenue per MWh$_e$ depend on the gold price at $Q = 20$, with $\mrm{LCON}^\mrm{elec}$ shown at both electricity prices.

\begin{figure}[tb!]
    \centering
    \begin{subfigure}[t]{\textwidth}
    \centering
    \includegraphics[width=0.99\textwidth]{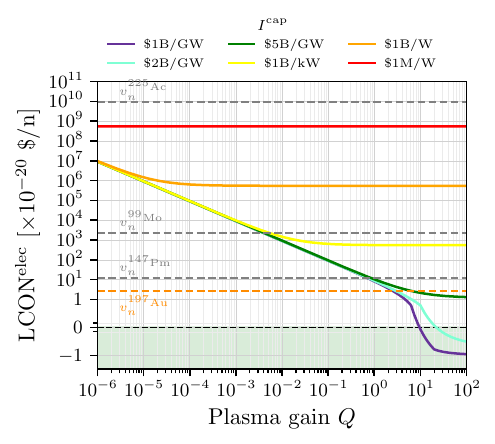}
    \end{subfigure}
    \caption{Levelized cost of a neutron for electricity only (LCON$^\mathrm{elec}$, solid) versus plasma gain for six capital intensities at $C_e = \$100$/MWh$_e$ and $\mathcal{A} = 1$. Dashed horizontal lines: value per neutron $v_\mathrm{n}$ for transmutation products. A product is profitable where its $v_\mathrm{n}$ line lies above the LCON$^\mathrm{elec}$ curve. Green shading: profitable region where LCON$^\mathrm{elec} < 0$ (electricity revenue alone exceeds all costs). Standard parameters from \cite{parisi2025isotope}.}
\label{fig:LCON}
\end{figure}

The LCON maps directly onto LCOE for net-electricity-producing plants ($\varepsilon_e > 0$, i.e.\ above engineering breakeven). The LCOE is the $C_e$ at which $\mrm{LCON}^\mrm{elec} = 0$. Since $c_\mrm{n}^\mrm{op}$ and $v_\mrm{n}^\mrm{elec}$ are both proportional to $C_e$, solving gives
\begin{equation}
    \mrm{LCOE} = \frac{c_\mrm{n}^\mrm{cap}}{\varepsilon_e}\cdot 3.6 \times 10^9 \;\; [\$/\mrm{MWh}_e],
    \label{eq:LCOE}
\end{equation}
where $\varepsilon_e \equiv (v_\mrm{n}^\mrm{elec} - c_\mrm{n}^\mrm{op})/\tilde{C}_e$ is the net electrical energy per source neutron (J/n). With electricity and isotope co-production, LCOE becomes
\begin{equation}
    \mrm{LCOE}^{\mrm{elec+prod}} = \frac{c_\mrm{n}^\mrm{cap} - \sum_i v_\mrm{n}^{\mrm{prod},i}}{\varepsilon_e}\cdot 3.6 \times 10^9.
    \label{eq:LCOE_cogen}
\end{equation}
\Cref{eq:LCOE,eq:LCOE_cogen} are valid only when $\varepsilon_e > 0$; below engineering breakeven the plant is a net electricity consumer and LCOE is not defined. This offset is powerful because transmutation increases revenue with relatively little additional capital. The fusion plant (plasma, magnets, blanket, and balance of plant) dominates $c_\mrm{n}^\mrm{cap}$ while feedstock loading into an existing blanket is a much smaller cost. Transmutation revenue therefore lower LCOE. At current gold prices, $v_\mrm{n}^\mrm{gold} \approx 2.4\, c_\mrm{n}^\mrm{cap}$: gold revenue alone exceeds the amortized capital cost per neutron, driving $\mrm{LCOE}^{\mrm{elec+prod}}$ negative.

For high-value radioisotopes such as \ce{^{99}Mo} ($v_\mathrm{n} \sim 10^{-17}$\,\$/n), the neutron value exceeds the LCON by orders of magnitude unless $Q \lesssim 10^{-2}$, so the binding constraint is the market size (\cref{eq:Pfus_market}), not the LCON. Recent work has shown a single ten megawatt D-T fusion device could fulfill global demand for many major medical radioisotopes \cite{parisi2025j}.

\Cref{fig:LCON} shows LCON$^\mathrm{elec}$ versus $Q$ for six capital intensities, with horizontal lines marking $v_\mathrm{n}$ for representative products. The intersection of a LCON curve with a product-value line gives the minimum gain for profitable transmutation of that product. Gold ($v_\mathrm{n} \approx \$2.6\cdot10^{-20}$/n) requires $Q \gtrsim 4$ at $\$2$\,B/GW, while \ce{^{99}Mo} ($v_\mathrm{n} \approx \$2.6\cdot10^{-17}$/n) is profitable at $Q \sim 10^{-3}$. At sufficiently high gain ($Q \gtrsim 10$ for $\$2$\,B/GW), LCON$^\mathrm{elec}$ becomes negative: electricity revenue alone exceeds all costs, and electricity-only fusion plants become viable.

\textit{Rate versus flux.} The effective LCON also depends on the neutron flux, sometimes strongly. Profitability requires that the feedstock be sufficiently cheap and/or consumed sufficiently quickly. In the low neutron flux limit the feedstock burns slowly and its inventory acts as a capital cost amortized over the plant lifetime~\cite{parisi2025isotope}. The amortized feedstock cost per neutron is
\begin{equation}
    c_\mathrm{n}^\mathrm{feed} = \frac{C_\mathrm{feed}\, M_\mathrm{feed}}{\dot{N}_\mathrm{n}\, T_\mathrm{year}\, \mathcal{A}\, S_\mathrm{disc}},
    \label{eq:cn_feed}
\end{equation}
where $C_\mathrm{feed}$ is the feedstock price per unit mass and $M_\mathrm{feed}$ the total inventory mass. The initial capital cost satisfies $c^\mathrm{cap}_\mathrm{n} = c^\mathrm{feed}_\mathrm{n} + c^\mathrm{plant}_\mathrm{n}$, where $c^\mathrm{plant}_\mathrm{n}$ describes all non-feedstock costs. In the thin-blanket limit \cite{parisi2025isotope}, $M_\mathrm{feed} \propto A$ and $\dot{N}_\mathrm{n} = \phi_\mathrm{n} A$, where $A$ is the blanket area and $\phi_\mathrm{n}$ the first-wall neutron flux (n\,cm$^{-2}$\,s$^{-1}$), so $c_\mathrm{n}^\mathrm{feed} \propto 1/(\phi_\mathrm{n}\,\mathcal{A})$. In LCON the feedstock loading is neutron-flux-dependent,
\begin{equation}
\begin{aligned}
\mathrm{LCON}^{\mrm{elec+prod}}(\phi_\mathrm{n}) = & c_\mathrm{n}^\mathrm{plant} + c_\mathrm{n}^\mathrm{feed}(\phi_\mathrm{n}) + c_\mathrm{n}^\mathrm{op} \\ & - v_\mathrm{n}^\mathrm{elec} - \sum_i v_\mathrm{n}^{\mathrm{prod},i}.
\end{aligned}
\label{eq:LCON_flux}
\end{equation}
Because $c_\mathrm{n}^\mathrm{feed} \propto 1/\phi_\mathrm{n}$, the LCON diverges at low flux; below $\phi_\mathrm{n}^\mathrm{min}$ the discounted product revenue can never repay the feedstock~\cite{parisi2025isotope} and $\mathrm{LCON} \to \infty$.

\begin{figure}[tb!]
    \centering
    \begin{subfigure}[t]{\textwidth}
    \centering
    \includegraphics[width=0.99\textwidth]{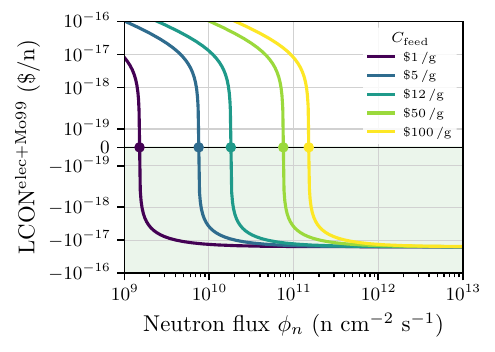}
    \end{subfigure}
    \caption{Levelized cost of a neutron for $\ce{^99Mo}$ production, LCON$^{\mathrm{elec}+\ce{^99Mo}}$. We used $Q=0.01$, $I^\mathrm{cap}= \$2$B/GW, $\mathcal{A} = 1$, $\sigma = 20$mb. Nominal cost of unenriched ruthenium: $C_\mathrm{feed}=\$12$/g. Green shading indicates the profitable region.}
\label{fig:LCON_Mo99_versus_flux}
\end{figure}

For expensive feedstocks such as \ce{^{226}Ra} ($\sim$\$1-50/$\mu$g), $c_\mathrm{n}^\mathrm{feed}$ can dominate $c_\mathrm{n}^\mathrm{cap}$, making high $\phi_\mathrm{n}$ essential for profitability. \Cref{fig:LCON_Mo99_versus_flux} illustrates this for \ce{^{99}Mo} production: at higher flux, feedstock costs are amortized faster, making each neutron more valuable for transmutation. Neutrons are not fungible - their value depends on the collective property $\phi_\mathrm{n}$.

\textit{The neutron cost gap.} How do current neutron sources compare with the targets in \Cref{fig:LCON}? \Cref{tab:gap} lists estimated LCON across existing and planned facilities. D-T neutron generators produce 14.1\,MeV neutrons at $\gtrsim\!\$10^{-13}$/n. The gap between today's practical high-energy neutron sources ($\sim\!\$10^{-13}$/n) and commercial fusion targets ($\sim\!\$10^{-20}$/n) spans roughly seven orders of magnitude. While this appears daunting, a large fraction of the gap is technological overhang: it reflects the low availability and subcommercial scale of current experiments, not fundamental limitations on the cost of confining a fusion plasma. Although the analysis in this Letter is primarily framed around magnetic confinement fusion (MCF), we include NIF as an inertial confinement reference point for completeness: despite achieving ignition~\cite{hurricane2024energy}, NIF has an LCON of $\sim\!\$10^{-10}$/n, estimated from the cumulative fusion yield of publicly reported shots~\cite{lasers_llnl_ignition} and total facility cost ($\sim$\$8\,B), owing to its high facility cost and very low repetition rate. JET, the most capable MCF D-T device ($Q \approx 0.33$~\cite{Keilhacker1999}), produced neutrons at $\sim\!\$10^{-12}$/n - comparable to D-T generators - because its LCON is dominated by the extremely low D-T availability ($\mathcal{A} \sim 10^{-6}$ for D-T operation) and high capital intensity of a research facility. In both cases, the high LCON reflects experimental operating conditions, not fundamental limitations on plasma performance. Operating at high availability in a commercial plant would reduce the LCON by several orders of magnitude.

\begin{table}[b]
\centering
\caption{Order-of-magnitude LCON and neutron flux for existing and projected neutron sources, including amortized capital and operating costs.}
\label{tab:gap}
\begin{ruledtabular}
\begin{tabular}{lcc}
Source & LCON & Fast Neutron Flux \\
 & (\$/neutron) & (n/cm${}^2$s) \\
\hline
NIF & $\sim\!10^{-10}$ & --- \\
JET & $\sim\!10^{-12}$ & $\sim\!10^{13}$$^\dagger$ \\
D-T generators & $\gtrsim\!10^{-13}$ & $\lesssim10^{10}$ \\
Volumetric neutron source & $\gtrsim\!10^{-19}$ & $\lesssim10^{14}$ \\
Commercial power plant & $\sim\!10^{-20}$ & $10^{13}$-$10^{14}$ \\
\multicolumn{3}{l}{\footnotesize $^\dagger$Instantaneous flux during a D-T pulse; duty-averaged flux is} \\
\multicolumn{3}{l}{\footnotesize much lower. VNS and commercial fluxes are steady-state.} \\
\end{tabular}
\end{ruledtabular}
\end{table}

To make this concrete: at $v_\mathrm{n}^\mathrm{gold}  = \$2.6\cdot10^{-20}$/n for gold (\Cref{tab:vn}), neutrons from a D-T generator ($\gtrsim\!\$10^{-13}$/n) cost at least $\sim\!10^7$ times more than the gold they produce. Only commercial-scale fusion at $Q \gtrsim 3$, lower CAPEX, and high availability brings the LCON below $v_\mathrm{n}^\mathrm{gold}$.

\textit{Closing the gap.} The seven-order-of-magnitude cost reduction from $\sim\!\$10^{-13}$/n to $\sim\!\$10^{-20}$/n breaks down into two qualitatively distinct stages.

\emph{The technological overhang ($\sim$5 orders of magnitude).} The dominant contributor to the high LCON of current fusion experiments is low availability. From \cref{eq:cn_cap}, $c_\mathrm{n}^\mrm{cap} \propto 1/\mathcal{A}$: a tokamak discharge lasting seconds to minutes, followed by hours or days of downtime, gives $\mathcal{A} \sim 10^{-5}$-$10^{-2}$. A commercial plant operating at $\mathcal{A} > 0.9$ amortizes capital over 2-4 orders of magnitude more neutrons at the same instantaneous fusion power. This alone accounts for the majority of the cost gap. Combined with increases in $Q$ from current experimental values ($Q \sim 10^{-2}$-$10^{-1}$) to near-term targets ($Q \sim 1$-$3$), which reduce $c_\mathrm{n}^\mathrm{op}$ by 1-2 orders of magnitude via \cref{eq:LCON_simple}, the LCON falls to $\sim\!\$10^{-18}$/n without requiring any reduction in capital intensity.

\emph{The last mile ($\sim$2 orders of magnitude).} Reaching $\sim\!\$10^{-20}$/n from $\sim\!\$10^{-18}$/n is hard. This requires simultaneously (i) increasing $Q$ from $<$1 to $\gtrsim$10-20 (or $\gtrsim 3$ if co-producing gold), which further reduces $c_\mathrm{n}^\mrm{op}$ until electricity revenue exceeds recirculating power costs, and (ii) reducing capital intensity from ITER-class costs ($\sim\!\$50$\,B/GW) to commercial targets ($\sim\!\$1$-$5$\,B/GW), dropping $c_\mathrm{n}^\mrm{cap}$ by $\sim$1-1.5 orders of magnitude. These improvements are more tightly coupled: high $Q$ enables smaller and cheaper heating systems, while lower CAPEX demands compact, high-field, or otherwise cost-optimized designs. Beyond $Q \sim 5$, alpha heating dominates and the marginal economic benefit per unit of $Q$ diminishes, but remains useful for reducing heating-system CAPEX and simplifying heat exhaust.

\begin{figure}[tb!]
    \centering
    \begin{subfigure}[t]{\textwidth}
    \centering
    \includegraphics[width=1.0\textwidth]{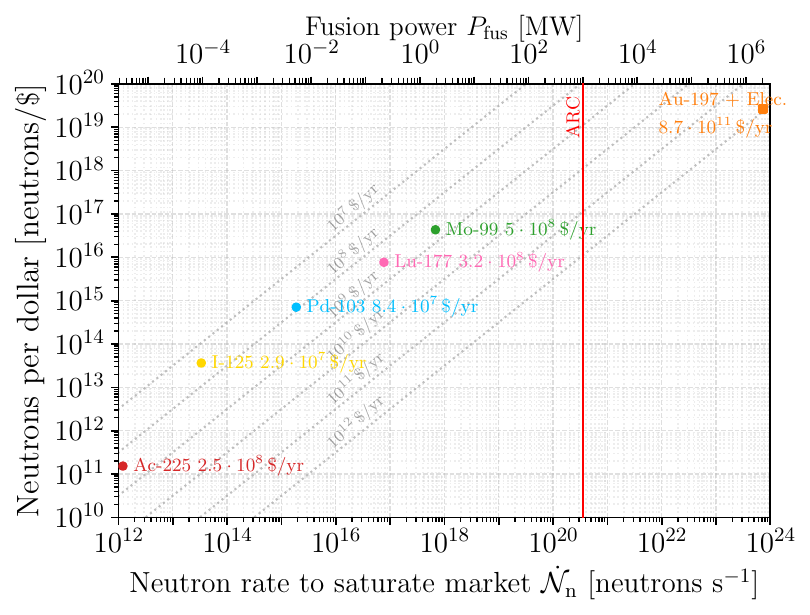}
    \end{subfigure}
    \caption{Neutrons per dollar versus neutron rate for entire market for a range of transmutation products. ARC \cite{Sorbom2015} vertical line assumes 100\% availability factor.}
\label{fig:value_per_neutron_rate}
\end{figure}

\textit{The neutron ladder.} Because neutron values span so many orders of magnitude (\Cref{tab:vn}), there is a natural staged development pathway. The most valuable products are profitable even when neutrons are very expensive, so early fusion systems can generate revenue well before reaching electricity-competitive performance. Some noteworthy transmutation products include: \ce{^{225}Ac} ($v_\mathrm{n} \sim 10^{-10}$\,\$/n), which is profitable with existing D-T sources, provided \ce{^{226}Ra} feedstock is sufficiently inexpensive and the neutron flux sufficiently high \cite{parisi2025j,parisi2025isotopemuon}; \ce{^{99}Mo} ($v_\mathrm{n} \sim 10^{-17}$\,\$/n), where a \qty{3}{MW} system with $Q \ll 1$ could supply global \ce{^{99}Mo} demand (\$0.7B/yr)~\cite{parisi2025j,parisi2025isotope}; \ce{^{147}Pm} ($v_\mathrm{n} \sim 10^{-19}$\,\$/n), which is viable at $Q \lesssim 1$, with a market supporting $\sim$\qty{100}{MW} of fusion capacity~\cite{parisi2025j,parisi2025isotope}; \ce{^{197}Au} ($v_\mathrm{n} \sim 10^{-20}$\,\$/n), viable at $Q \gtrsim 3$, with a $\sim$\$700\,B/yr market supporting $\sim$2\,TW of capacity~\cite{rutkowski2025scalable,parisi2025isotope}; and electricity ($v_\mathrm{n} \sim 10^{-20}$\,\$/n), which requires $Q \gtrsim 7$-$10$ for standalone viability at $I^\mathrm{cap}=$1-2\,B/GW.

\noindent Each rung generates revenue and operational experience that funds the next. Fusion does not need to reach electricity-competitive performance before generating significant revenue: the most valuable products per neutron are accessible with modest devices, and their revenue can help fund higher-performance systems that unlock progressively larger markets.

\textit{Market size and fusion capacity.} The total fusion power that a transmutation market can support is determined by market size $S_\mrm{market}$ and the value per neutron~\cite{parisi2025isotope},
\begin{equation}
    P_\mrm{fus}^\mrm{market} \equiv \dot{\mathcal{N}}_\mathrm{n}\, E_\mrm{fus} = \frac{S_\mrm{market}\, E_\mrm{fus}}{v_\mathrm{n}\, T_\mrm{year}\, \mathcal{A}},
    \label{eq:Pfus_market}
\end{equation}
where $\dot{\mathcal{N}}_\mathrm{n}$ is the corresponding D-T neutron rate for the entire market. High-$v_\mathrm{n}$ products require little fusion power to saturate their markets, while low-$v_\mathrm{n}$ products can support terawatts of capacity. The \ce{^{99}Mo} market ($\sim$\$0.7B/yr) saturates at only $\sim$\qty{3}{MW} of fusion power, ideal for early MW-class machines. The gold market ($\sim$\$700\,B/yr) can support $\sim$2\,TW~\cite{parisi2025isotope,rutkowski2025scalable}, comparable to what is needed for fusion to contribute meaningfully to global energy supply. Even 10\% of the gold market would support $\sim$250\,GW at $Q \sim 2$-$3$, well before electricity-only operation is required. \Cref{fig:value_per_neutron_rate} shows this trade-off. Early markets only require fusion deployment at the watt to megawatt scale.

Fusion is uniquely suited to large-scale transmutation because a plant can simultaneously sell electricity and transmutation products, using electricity revenue to offset operating costs. This is the origin of the negative $c_\mathrm{n}^\mrm{op}$ at high gain (\cref{eq:cn_op}): the electricity credit makes each neutron progressively cheaper as plasma performance improves. For smaller sub-MW machines, however, electricity generation is likely nonviable due to high CAPEX for generation equipment.

\textit{Discussion.} High-energy fusion neutrons, often viewed as a liability of D-T reactions, are a source of significant economic and social value. The LCON provides a simple metric for whether a fusion neutron is worth making. Today's sources are $\sim$7 orders of magnitude too expensive for commercial fusion electricity generation. Most of this gap ($\sim$5 orders of magnitude) is technological overhang that will be closed by operating existing confinement concepts at high availability and modest gain; the remaining $\sim$2 orders of magnitude - the last mile - requires higher $Q$ and lower capital intensity. But the gap need not be closed all at once. Because neutron values span ten orders of magnitude - from $\sim\!\$10^{-20}$/n for electricity to $\sim\!\$10^{-10}$/n for \ce{^{225}Ac} - each improvement in fusion performance unlocks new transmutation products with progressively larger markets. The neutron ladder offers a revenue-positive path from small transmutation-only machines to terawatt-scale power plants co-producing electricity and gold. Long term, driving down the neutron cost by many orders of magnitude will enable applications beyond those described here.

\bibliography{Master_EverythingPlasmaBib} %

\begin{thebibliography}{23}%
\makeatletter
\providecommand \@ifxundefined [1]{%
 \@ifx{#1\undefined}
}%
\providecommand \@ifnum [1]{%
 \ifnum #1\expandafter \@firstoftwo
 \else \expandafter \@secondoftwo
 \fi
}%
\providecommand \@ifx [1]{%
 \ifx #1\expandafter \@firstoftwo
 \else \expandafter \@secondoftwo
 \fi
}%
\providecommand \natexlab [1]{#1}%
\providecommand \enquote  [1]{``#1''}%
\providecommand \bibnamefont  [1]{#1}%
\providecommand \bibfnamefont [1]{#1}%
\providecommand \citenamefont [1]{#1}%
\providecommand \href@noop [0]{\@secondoftwo}%
\providecommand \href [0]{\begingroup \@sanitize@url \@href}%
\providecommand \@href[1]{\@@startlink{#1}\@@href}%
\providecommand \@@href[1]{\endgroup#1\@@endlink}%
\providecommand \@sanitize@url [0]{\catcode `\\12\catcode `\$12\catcode
  `\&12\catcode `\#12\catcode `\^12\catcode `\_12\catcode `\%12\relax}%
\providecommand \@@startlink[1]{}%
\providecommand \@@endlink[0]{}%
\providecommand \url  [0]{\begingroup\@sanitize@url \@url }%
\providecommand \@url [1]{\endgroup\@href {#1}{\urlprefix }}%
\providecommand \urlprefix  [0]{URL }%
\providecommand \Eprint [0]{\href }%
\providecommand \doibase [0]{https://doi.org/}%
\providecommand \selectlanguage [1]{}%
\providecommand \bibinfo  [0]{\@secondoftwo}%
\providecommand \bibfield  [0]{\@secondoftwo}%
\providecommand \translation [1]{[#1]}%
\providecommand \BibitemOpen [0]{}%
\providecommand \bibitemStop [0]{}%
\providecommand \bibitemNoStop [0]{.\EOS\space}%
\providecommand \EOS [0]{\spacefactor3000\relax}%
\providecommand \BibitemShut  [1]{\csname bibitem#1\endcsname}%
\let\auto@bib@innerbib\@empty
\bibitem [{\citenamefont {Nordhaus}(1996)}]{nordhaus1996real}%
  \BibitemOpen
  \bibfield  {author} {\bibinfo {author} {\bibfnamefont {W.~D.}\ \bibnamefont
  {Nordhaus}},\ }\bibfield  {title} {\bibinfo {title} {Do real-output and
  real-wage measures capture reality? the history of lighting suggests not},\
  }in\ \href@noop {} {\emph {\bibinfo {booktitle} {The economics of new
  goods}}}\ (\bibinfo  {publisher} {University of Chicago Press},\ \bibinfo
  {year} {1996})\ pp.\ \bibinfo {pages} {27--70}\BibitemShut {NoStop}%
\bibitem [{\citenamefont {Smil}(2018)}]{smil2018energy}%
  \BibitemOpen
  \bibfield  {author} {\bibinfo {author} {\bibfnamefont {V.}~\bibnamefont
  {Smil}},\ }\href@noop {} {\emph {\bibinfo {title} {Energy and civilization: a
  history}}}\ (\bibinfo  {publisher} {MIT press},\ \bibinfo {year}
  {2018})\BibitemShut {NoStop}%
\bibitem [{\citenamefont {Potter}(2025)}]{potter2025origins}%
  \BibitemOpen
  \bibfield  {author} {\bibinfo {author} {\bibfnamefont {B.}~\bibnamefont
  {Potter}},\ }\href@noop {} {\emph {\bibinfo {title} {The Origins of
  Efficiency}}}\ (\bibinfo  {publisher} {Stripe Press},\ \bibinfo {year}
  {2025})\BibitemShut {NoStop}%
\bibitem [{\citenamefont {Wurzel}\ and\ \citenamefont
  {Hsu}(2022)}]{Wurzel2022}%
  \BibitemOpen
  \bibfield  {author} {\bibinfo {author} {\bibfnamefont {S.~E.}\ \bibnamefont
  {Wurzel}}\ and\ \bibinfo {author} {\bibfnamefont {S.~C.}\ \bibnamefont
  {Hsu}},\ }\bibfield  {title} {\bibinfo {title} {Progress toward fusion energy
  breakeven and gain as measured against the lawson criterion},\ }\bibfield
  {journal} {\bibinfo  {journal} {Physics of Plasmas}\ }\textbf {\bibinfo
  {volume} {29}},\ \href {https://doi.org/10.1063/5.0083990}
  {10.1063/5.0083990} (\bibinfo {year} {2022})\BibitemShut {NoStop}%
\bibitem [{\citenamefont {Engholm}\ \emph {et~al.}(1986)\citenamefont
  {Engholm}, \citenamefont {Cheng},\ and\ \citenamefont
  {Schultz}}]{engholm1986radioisotope}%
  \BibitemOpen
  \bibfield  {author} {\bibinfo {author} {\bibfnamefont {B.~A.}\ \bibnamefont
  {Engholm}}, \bibinfo {author} {\bibfnamefont {E.~T.}\ \bibnamefont {Cheng}},\
  and\ \bibinfo {author} {\bibfnamefont {K.~R.}\ \bibnamefont {Schultz}},\
  }\bibfield  {title} {\bibinfo {title} {Radioisotope production in fusion
  reactors},\ }\href@noop {} {\bibfield  {journal} {\bibinfo  {journal} {Fusion
  technology}\ }\textbf {\bibinfo {volume} {10}},\ \bibinfo {pages} {1290}
  (\bibinfo {year} {1986})}\BibitemShut {NoStop}%
\bibitem [{\citenamefont {Rutkowski}\ \emph {et~al.}(2025)\citenamefont
  {Rutkowski}, \citenamefont {Harter},\ and\ \citenamefont
  {Parisi}}]{rutkowski2025scalable}%
  \BibitemOpen
  \bibfield  {author} {\bibinfo {author} {\bibfnamefont {A.}~\bibnamefont
  {Rutkowski}}, \bibinfo {author} {\bibfnamefont {J.}~\bibnamefont {Harter}},\
  and\ \bibinfo {author} {\bibfnamefont {J.}~\bibnamefont {Parisi}},\
  }\bibfield  {title} {\bibinfo {title} {Scalable chrysopoeia via $(n, 2n) $
  reactions driven by deuterium-tritium fusion neutrons},\ }\href@noop {}
  {\bibfield  {journal} {\bibinfo  {journal} {arXiv preprint arXiv:2507.13461}\
  } (\bibinfo {year} {2025})}\BibitemShut {NoStop}%
\bibitem [{\citenamefont {Parisi}\ \emph
  {et~al.}(2025{\natexlab{a}})\citenamefont {Parisi}, \citenamefont {Schwartz},
  \citenamefont {Wurzel}, \citenamefont {Rutkowski},\ and\ \citenamefont
  {Harter}}]{parisi2025isotope}%
  \BibitemOpen
  \bibfield  {author} {\bibinfo {author} {\bibfnamefont {J.~F.}\ \bibnamefont
  {Parisi}}, \bibinfo {author} {\bibfnamefont {J.~A.}\ \bibnamefont
  {Schwartz}}, \bibinfo {author} {\bibfnamefont {S.~E.}\ \bibnamefont
  {Wurzel}}, \bibinfo {author} {\bibfnamefont {A.}~\bibnamefont {Rutkowski}},\
  and\ \bibinfo {author} {\bibfnamefont {J.}~\bibnamefont {Harter}},\
  }\bibfield  {title} {\bibinfo {title} {Isotope production in fusion
  systems},\ }\href@noop {} {\bibfield  {journal} {\bibinfo  {journal} {arXiv
  preprint arXiv:2512.09242}\ } (\bibinfo {year}
  {2025}{\natexlab{a}})}\BibitemShut {NoStop}%
\bibitem [{\citenamefont {Parisi}\ \emph
  {et~al.}(2025{\natexlab{b}})\citenamefont {Parisi}, \citenamefont
  {Rutkowski}, \citenamefont {Harter}, \citenamefont {Schwartz},\ and\
  \citenamefont {Chen}}]{parisi2025j}%
  \BibitemOpen
  \bibfield  {author} {\bibinfo {author} {\bibfnamefont {J.~F.}\ \bibnamefont
  {Parisi}}, \bibinfo {author} {\bibfnamefont {A.}~\bibnamefont {Rutkowski}},
  \bibinfo {author} {\bibfnamefont {J.}~\bibnamefont {Harter}}, \bibinfo
  {author} {\bibfnamefont {J.~A.}\ \bibnamefont {Schwartz}},\ and\ \bibinfo
  {author} {\bibfnamefont {S.}~\bibnamefont {Chen}},\ }\href
  {https://arxiv.org/abs/2511.02814} {\bibinfo {title} {Production of
  high-specific-activity radioisotopes using high-energy fusion neutrons}}
  (\bibinfo {year} {2025}{\natexlab{b}}),\ \Eprint
  {https://arxiv.org/abs/2511.02814} {arXiv:2511.02814 [nucl-ex]} \BibitemShut
  {NoStop}%
\bibitem [{\citenamefont {Evitts}\ \emph {et~al.}()\citenamefont {Evitts},
  \citenamefont {Miller}, \citenamefont {{Da Pieve}}, \citenamefont {Turner},\
  and\ \citenamefont {Borini}}]{evitts2025theoretical}%
  \BibitemOpen
  \bibfield  {author} {\bibinfo {author} {\bibfnamefont {L.~J.}\ \bibnamefont
  {Evitts}}, \bibinfo {author} {\bibfnamefont {P.~W.}\ \bibnamefont {Miller}},
  \bibinfo {author} {\bibfnamefont {C.}~\bibnamefont {{Da Pieve}}}, \bibinfo
  {author} {\bibfnamefont {A.}~\bibnamefont {Turner}},\ and\ \bibinfo {author}
  {\bibfnamefont {S.}~\bibnamefont {Borini}},\ }\bibfield  {title} {\bibinfo
  {title} {Theoretical novel medical isotope production with deuterium-tritium
  fusion technology},\ }\href
  {https://doi.org/https://doi.org/10.1016/j.apradiso.2025.112163} {\bibfield
  {journal} {\bibinfo  {journal} {Applied Radiation and Isotopes}\ }\textbf
  {\bibinfo {volume} {226}},\ \bibinfo {pages} {112163}}\BibitemShut {NoStop}%
\bibitem [{\citenamefont {Parisi}\ and\ \citenamefont
  {Rutkowski}(2025)}]{parisi2025isotopemuon}%
  \BibitemOpen
  \bibfield  {author} {\bibinfo {author} {\bibfnamefont {J.~F.}\ \bibnamefont
  {Parisi}}\ and\ \bibinfo {author} {\bibfnamefont {A.}~\bibnamefont
  {Rutkowski}},\ }\bibfield  {title} {\bibinfo {title} {Isotope production in
  muon-catalyzed-fusion systems},\ }\href@noop {} {\bibfield  {journal}
  {\bibinfo  {journal} {arXiv preprint arXiv:2511.20951}\ } (\bibinfo {year}
  {2025})}\BibitemShut {NoStop}%
\bibitem [{\citenamefont {Branker}\ \emph {et~al.}(2011)\citenamefont
  {Branker}, \citenamefont {Pathak},\ and\ \citenamefont
  {Pearce}}]{branker2011review}%
  \BibitemOpen
  \bibfield  {author} {\bibinfo {author} {\bibfnamefont {K.}~\bibnamefont
  {Branker}}, \bibinfo {author} {\bibfnamefont {M.}~\bibnamefont {Pathak}},\
  and\ \bibinfo {author} {\bibfnamefont {J.~M.}\ \bibnamefont {Pearce}},\
  }\bibfield  {title} {\bibinfo {title} {A review of solar photovoltaic
  levelized cost of electricity},\ }\href@noop {} {\bibfield  {journal}
  {\bibinfo  {journal} {Renewable and sustainable energy reviews}\ }\textbf
  {\bibinfo {volume} {15}},\ \bibinfo {pages} {4470} (\bibinfo {year}
  {2011})}\BibitemShut {NoStop}%
\bibitem [{\citenamefont {Kost}\ \emph {et~al.}(2013)\citenamefont {Kost},
  \citenamefont {Mayer}, \citenamefont {Thomsen}, \citenamefont {Hartmann},
  \citenamefont {Senkpiel}, \citenamefont {Philipps}, \citenamefont {Nold},
  \citenamefont {Lude}, \citenamefont {Saad},\ and\ \citenamefont
  {Schlegl}}]{kost2013levelized}%
  \BibitemOpen
  \bibfield  {author} {\bibinfo {author} {\bibfnamefont {C.}~\bibnamefont
  {Kost}}, \bibinfo {author} {\bibfnamefont {J.~N.}\ \bibnamefont {Mayer}},
  \bibinfo {author} {\bibfnamefont {J.}~\bibnamefont {Thomsen}}, \bibinfo
  {author} {\bibfnamefont {N.}~\bibnamefont {Hartmann}}, \bibinfo {author}
  {\bibfnamefont {C.}~\bibnamefont {Senkpiel}}, \bibinfo {author}
  {\bibfnamefont {S.}~\bibnamefont {Philipps}}, \bibinfo {author}
  {\bibfnamefont {S.}~\bibnamefont {Nold}}, \bibinfo {author} {\bibfnamefont
  {S.}~\bibnamefont {Lude}}, \bibinfo {author} {\bibfnamefont {N.}~\bibnamefont
  {Saad}},\ and\ \bibinfo {author} {\bibfnamefont {T.}~\bibnamefont
  {Schlegl}},\ }\bibfield  {title} {\bibinfo {title} {Levelized cost of
  electricity-renewable energy technologies},\ }\href@noop {} {\  (\bibinfo
  {year} {2013})}\BibitemShut {NoStop}%
\bibitem [{\citenamefont {Obi}\ \emph {et~al.}(2017)\citenamefont {Obi},
  \citenamefont {Jensen}, \citenamefont {Ferris},\ and\ \citenamefont
  {Bass}}]{obi2017calculation}%
  \BibitemOpen
  \bibfield  {author} {\bibinfo {author} {\bibfnamefont {M.}~\bibnamefont
  {Obi}}, \bibinfo {author} {\bibfnamefont {S.~M.}\ \bibnamefont {Jensen}},
  \bibinfo {author} {\bibfnamefont {J.~B.}\ \bibnamefont {Ferris}},\ and\
  \bibinfo {author} {\bibfnamefont {R.~B.}\ \bibnamefont {Bass}},\ }\bibfield
  {title} {\bibinfo {title} {Calculation of levelized costs of electricity for
  various electrical energy storage systems},\ }\href@noop {} {\bibfield
  {journal} {\bibinfo  {journal} {Renewable and Sustainable Energy Reviews}\
  }\textbf {\bibinfo {volume} {67}},\ \bibinfo {pages} {908} (\bibinfo {year}
  {2017})}\BibitemShut {NoStop}%
\bibitem [{\citenamefont {Hansen}(2019)}]{hansen2019decision}%
  \BibitemOpen
  \bibfield  {author} {\bibinfo {author} {\bibfnamefont {K.}~\bibnamefont
  {Hansen}},\ }\bibfield  {title} {\bibinfo {title} {Decision-making based on
  energy costs: Comparing levelized cost of energy and energy system costs},\
  }\href@noop {} {\bibfield  {journal} {\bibinfo  {journal} {Energy Strategy
  Reviews}\ }\textbf {\bibinfo {volume} {24}},\ \bibinfo {pages} {68} (\bibinfo
  {year} {2019})}\BibitemShut {NoStop}%
\bibitem [{\citenamefont {Sens}\ \emph {et~al.}(2022)\citenamefont {Sens},
  \citenamefont {Neuling},\ and\ \citenamefont
  {Kaltschmitt}}]{sens2022capital}%
  \BibitemOpen
  \bibfield  {author} {\bibinfo {author} {\bibfnamefont {L.}~\bibnamefont
  {Sens}}, \bibinfo {author} {\bibfnamefont {U.}~\bibnamefont {Neuling}},\ and\
  \bibinfo {author} {\bibfnamefont {M.}~\bibnamefont {Kaltschmitt}},\
  }\bibfield  {title} {\bibinfo {title} {Capital expenditure and levelized cost
  of electricity of photovoltaic plants and wind turbines--development by
  2050},\ }\href@noop {} {\bibfield  {journal} {\bibinfo  {journal} {Renewable
  Energy}\ }\textbf {\bibinfo {volume} {185}},\ \bibinfo {pages} {525}
  (\bibinfo {year} {2022})}\BibitemShut {NoStop}%
\bibitem [{\citenamefont {Schwartz}\ \emph {et~al.}(2023)\citenamefont
  {Schwartz}, \citenamefont {Ricks}, \citenamefont {Kolemen},\ and\
  \citenamefont {Jenkins}}]{schwartz2023value}%
  \BibitemOpen
  \bibfield  {author} {\bibinfo {author} {\bibfnamefont {J.~A.}\ \bibnamefont
  {Schwartz}}, \bibinfo {author} {\bibfnamefont {W.}~\bibnamefont {Ricks}},
  \bibinfo {author} {\bibfnamefont {E.}~\bibnamefont {Kolemen}},\ and\ \bibinfo
  {author} {\bibfnamefont {J.~D.}\ \bibnamefont {Jenkins}},\ }\bibfield
  {title} {\bibinfo {title} {The value of fusion energy to a decarbonized
  united states electric grid},\ }\href@noop {} {\bibfield  {journal} {\bibinfo
   {journal} {Joule}\ }\textbf {\bibinfo {volume} {7}},\ \bibinfo {pages} {675}
  (\bibinfo {year} {2023})}\BibitemShut {NoStop}%
\bibitem [{\citenamefont {Sawan}\ and\ \citenamefont
  {Abdou}(2006)}]{sawan_physics_2006}%
  \BibitemOpen
  \bibfield  {author} {\bibinfo {author} {\bibfnamefont {M.}~\bibnamefont
  {Sawan}}\ and\ \bibinfo {author} {\bibfnamefont {M.}~\bibnamefont {Abdou}},\
  }\bibfield  {title} {\bibinfo {title} {Physics and technology conditions for
  attaining tritium self-sufficiency for the {DT} fuel cycle},\ }\href
  {https://doi.org/10.1016/j.fusengdes.2005.07.035} {\bibfield  {journal}
  {\bibinfo  {journal} {Fusion Engineering and Design}\ }\textbf {\bibinfo
  {volume} {81}},\ \bibinfo {pages} {1131} (\bibinfo {year}
  {2006})}\BibitemShut {NoStop}%
\bibitem [{\citenamefont {Abdou}\ \emph {et~al.}(2021)\citenamefont {Abdou},
  \citenamefont {Riva}, \citenamefont {Ying}, \citenamefont {Day},
  \citenamefont {Loarte}, \citenamefont {Baylor}, \citenamefont {Humrickhouse},
  \citenamefont {Fuerst},\ and\ \citenamefont {Cho}}]{Abdou2021}%
  \BibitemOpen
  \bibfield  {author} {\bibinfo {author} {\bibfnamefont {M.}~\bibnamefont
  {Abdou}}, \bibinfo {author} {\bibfnamefont {M.}~\bibnamefont {Riva}},
  \bibinfo {author} {\bibfnamefont {A.}~\bibnamefont {Ying}}, \bibinfo {author}
  {\bibfnamefont {C.}~\bibnamefont {Day}}, \bibinfo {author} {\bibfnamefont
  {A.}~\bibnamefont {Loarte}}, \bibinfo {author} {\bibfnamefont {L.~R.}\
  \bibnamefont {Baylor}}, \bibinfo {author} {\bibfnamefont {P.}~\bibnamefont
  {Humrickhouse}}, \bibinfo {author} {\bibfnamefont {T.~F.}\ \bibnamefont
  {Fuerst}},\ and\ \bibinfo {author} {\bibfnamefont {S.}~\bibnamefont {Cho}},\
  }\bibfield  {title} {\bibinfo {title} {Physics and technology considerations
  for the deuterium-tritium fuel cycle and conditions for tritium fuel self
  sufficiency},\ }\bibfield  {journal} {\bibinfo  {journal} {Nuclear Fusion}\
  }\textbf {\bibinfo {volume} {61}},\ \href
  {https://doi.org/10.1088/1741-4326/abbf35} {10.1088/1741-4326/abbf35}
  (\bibinfo {year} {2021})\BibitemShut {NoStop}%
\bibitem [{\citenamefont {Meschini}\ \emph {et~al.}(2023)\citenamefont
  {Meschini}, \citenamefont {Ferry}, \citenamefont {Delaporte-Mathurin},\ and\
  \citenamefont {Whyte}}]{Meschini2023}%
  \BibitemOpen
  \bibfield  {author} {\bibinfo {author} {\bibfnamefont {S.}~\bibnamefont
  {Meschini}}, \bibinfo {author} {\bibfnamefont {S.~E.}\ \bibnamefont {Ferry}},
  \bibinfo {author} {\bibfnamefont {R.}~\bibnamefont {Delaporte-Mathurin}},\
  and\ \bibinfo {author} {\bibfnamefont {D.~G.}\ \bibnamefont {Whyte}},\
  }\bibfield  {title} {\bibinfo {title} {Modeling and analysis of the tritium
  fuel cycle for arc- and step-class d-t fusion power plants},\ }\bibfield
  {journal} {\bibinfo  {journal} {Nuclear Fusion}\ }\textbf {\bibinfo {volume}
  {63}},\ \href {https://doi.org/10.1088/1741-4326/acf3fc}
  {10.1088/1741-4326/acf3fc} (\bibinfo {year} {2023})\BibitemShut {NoStop}%
\bibitem [{\citenamefont {Hurricane}\ \emph {et~al.}(2024)\citenamefont
  {Hurricane}, \citenamefont {Callahan}, \citenamefont {Casey}, \citenamefont
  {Christopherson}, \citenamefont {Kritcher}, \citenamefont {Landen},
  \citenamefont {Maclaren}, \citenamefont {Nora}, \citenamefont {Patel},
  \citenamefont {Ralph} \emph {et~al.}}]{hurricane2024energy}%
  \BibitemOpen
  \bibfield  {author} {\bibinfo {author} {\bibfnamefont {O.~A.}\ \bibnamefont
  {Hurricane}}, \bibinfo {author} {\bibfnamefont {D.}~\bibnamefont {Callahan}},
  \bibinfo {author} {\bibfnamefont {D.}~\bibnamefont {Casey}}, \bibinfo
  {author} {\bibfnamefont {A.}~\bibnamefont {Christopherson}}, \bibinfo
  {author} {\bibfnamefont {A.}~\bibnamefont {Kritcher}}, \bibinfo {author}
  {\bibfnamefont {O.}~\bibnamefont {Landen}}, \bibinfo {author} {\bibfnamefont
  {S.}~\bibnamefont {Maclaren}}, \bibinfo {author} {\bibfnamefont
  {R.}~\bibnamefont {Nora}}, \bibinfo {author} {\bibfnamefont {P.}~\bibnamefont
  {Patel}}, \bibinfo {author} {\bibfnamefont {J.}~\bibnamefont {Ralph}}, \emph
  {et~al.},\ }\bibfield  {title} {\bibinfo {title} {Energy principles of
  scientific breakeven in an inertial fusion experiment},\ }\href@noop {}
  {\bibfield  {journal} {\bibinfo  {journal} {Physical Review Letters}\
  }\textbf {\bibinfo {volume} {132}},\ \bibinfo {pages} {065103} (\bibinfo
  {year} {2024})}\BibitemShut {NoStop}%
\bibitem [{\citenamefont {{Lawrence Livermore National
  Laboratory}}(2025)}]{lasers_llnl_ignition}%
  \BibitemOpen
  \bibfield  {author} {\bibinfo {author} {\bibnamefont {{Lawrence Livermore
  National Laboratory}}},\ }\href
  {https://lasers.llnl.gov/science/achieving-fusion-ignition} {\bibinfo {title}
  {Achieving fusion ignition}} (\bibinfo {year} {2025}),\ \bibinfo {note}
  {accessed February 2026}\BibitemShut {NoStop}%
\bibitem [{\citenamefont {Keilhacker}\ \emph {et~al.}(1999)\citenamefont
  {Keilhacker}, \citenamefont {Gibson}, \citenamefont {Gormezano},
  \citenamefont {Lomas}, \citenamefont {Thomas}, \citenamefont {Watkins},
  \citenamefont {Andrew}, \citenamefont {Balet}, \citenamefont {Borba},
  \citenamefont {Challis}, \citenamefont {Coffey}, \citenamefont {Cottrell},
  \citenamefont {Esch}, \citenamefont {Deliyanakis}, \citenamefont {Fasoli},
  \citenamefont {Gowers}, \citenamefont {Guo}, \citenamefont {Huysmans},
  \citenamefont {Jones}, \citenamefont {Kerner}, \citenamefont {K{\"{o,}}nig},
  \citenamefont {Loughlin}, \citenamefont {Maas}, \citenamefont {Marcus},
  \citenamefont {Nave}, \citenamefont {Rimini}, \citenamefont {Sadler},
  \citenamefont {Sharapov}, \citenamefont {Sips}, \citenamefont {Smeulders},
  \citenamefont {S{\"{o}}ldner}, \citenamefont {Taroni}, \citenamefont
  {Tubbing}, \citenamefont {von Hellermann}, \citenamefont {Ward},\ and\
  \citenamefont {Team}}]{Keilhacker1999}%
  \BibitemOpen
  \bibfield  {author} {\bibinfo {author} {\bibfnamefont {M.}~\bibnamefont
  {Keilhacker}}, \bibinfo {author} {\bibfnamefont {A.}~\bibnamefont {Gibson}},
  \bibinfo {author} {\bibfnamefont {C.}~\bibnamefont {Gormezano}}, \bibinfo
  {author} {\bibfnamefont {P.~J.}\ \bibnamefont {Lomas}}, \bibinfo {author}
  {\bibfnamefont {P.~R.}\ \bibnamefont {Thomas}}, \bibinfo {author}
  {\bibfnamefont {M.~L.}\ \bibnamefont {Watkins}}, \bibinfo {author}
  {\bibfnamefont {P.}~\bibnamefont {Andrew}}, \bibinfo {author} {\bibfnamefont
  {B.}~\bibnamefont {Balet}}, \bibinfo {author} {\bibfnamefont
  {D.}~\bibnamefont {Borba}}, \bibinfo {author} {\bibfnamefont {C.~D.}\
  \bibnamefont {Challis}}, \bibinfo {author} {\bibfnamefont {I.}~\bibnamefont
  {Coffey}}, \bibinfo {author} {\bibfnamefont {G.~A.}\ \bibnamefont
  {Cottrell}}, \bibinfo {author} {\bibfnamefont {H.~P. L.~D.}\ \bibnamefont
  {Esch}}, \bibinfo {author} {\bibfnamefont {N.}~\bibnamefont {Deliyanakis}},
  \bibinfo {author} {\bibfnamefont {A.}~\bibnamefont {Fasoli}}, \bibinfo
  {author} {\bibfnamefont {C.~W.}\ \bibnamefont {Gowers}}, \bibinfo {author}
  {\bibfnamefont {H.~Y.}\ \bibnamefont {Guo}}, \bibinfo {author} {\bibfnamefont
  {G.~T.~A.}\ \bibnamefont {Huysmans}}, \bibinfo {author} {\bibfnamefont
  {T.~T.~C.}\ \bibnamefont {Jones}}, \bibinfo {author} {\bibfnamefont
  {W.}~\bibnamefont {Kerner}}, \bibinfo {author} {\bibfnamefont
  {R.}~\bibnamefont {K{\"{o,}}nig}}, \bibinfo {author} {\bibfnamefont
  {M.}~\bibnamefont {Loughlin}}, \bibinfo {author} {\bibfnamefont
  {A.}~\bibnamefont {Maas}}, \bibinfo {author} {\bibfnamefont {F.}~\bibnamefont
  {Marcus}}, \bibinfo {author} {\bibfnamefont {M.}~\bibnamefont {Nave}},
  \bibinfo {author} {\bibfnamefont {F.}~\bibnamefont {Rimini}}, \bibinfo
  {author} {\bibfnamefont {G.}~\bibnamefont {Sadler}}, \bibinfo {author}
  {\bibfnamefont {S.}~\bibnamefont {Sharapov}}, \bibinfo {author}
  {\bibfnamefont {G.}~\bibnamefont {Sips}}, \bibinfo {author} {\bibfnamefont
  {P.}~\bibnamefont {Smeulders}}, \bibinfo {author} {\bibfnamefont
  {F.}~\bibnamefont {S{\"{o}}ldner}}, \bibinfo {author} {\bibfnamefont
  {A.}~\bibnamefont {Taroni}}, \bibinfo {author} {\bibfnamefont
  {B.}~\bibnamefont {Tubbing}}, \bibinfo {author} {\bibfnamefont
  {M.}~\bibnamefont {von Hellermann}}, \bibinfo {author} {\bibfnamefont
  {D.}~\bibnamefont {Ward}},\ and\ \bibinfo {author} {\bibfnamefont
  {J.}~\bibnamefont {Team}},\ }\bibfield  {title} {\bibinfo {title} {{High
  fusion performance from deuterium-tritium plasmas in JET}},\ }\href@noop {}
  {\bibfield  {journal} {\bibinfo  {journal} {Nuclear Fusion}\ }\textbf
  {\bibinfo {volume} {39}},\ \bibinfo {pages} {209} (\bibinfo {year}
  {1999})}\BibitemShut {NoStop}%
\bibitem [{\citenamefont {Sorbom}\ \emph {et~al.}(2015)\citenamefont {Sorbom},
  \citenamefont {Ball}, \citenamefont {Palmer}, \citenamefont {Mangiarotti},
  \citenamefont {Sierchio}, \citenamefont {Bonoli}, \citenamefont {Kasten},
  \citenamefont {Sutherland}, \citenamefont {Barnard}, \citenamefont
  {Haakonsen}, \citenamefont {Goh}, \citenamefont {Sung},\ and\ \citenamefont
  {Whyte}}]{Sorbom2015}%
  \BibitemOpen
  \bibfield  {author} {\bibinfo {author} {\bibfnamefont {B.~N.}\ \bibnamefont
  {Sorbom}}, \bibinfo {author} {\bibfnamefont {J.}~\bibnamefont {Ball}},
  \bibinfo {author} {\bibfnamefont {T.~R.}\ \bibnamefont {Palmer}}, \bibinfo
  {author} {\bibfnamefont {F.~J.}\ \bibnamefont {Mangiarotti}}, \bibinfo
  {author} {\bibfnamefont {J.~M.}\ \bibnamefont {Sierchio}}, \bibinfo {author}
  {\bibfnamefont {P.}~\bibnamefont {Bonoli}}, \bibinfo {author} {\bibfnamefont
  {C.}~\bibnamefont {Kasten}}, \bibinfo {author} {\bibfnamefont {D.~A.}\
  \bibnamefont {Sutherland}}, \bibinfo {author} {\bibfnamefont {H.~S.}\
  \bibnamefont {Barnard}}, \bibinfo {author} {\bibfnamefont {C.~B.}\
  \bibnamefont {Haakonsen}}, \bibinfo {author} {\bibfnamefont {J.}~\bibnamefont
  {Goh}}, \bibinfo {author} {\bibfnamefont {C.}~\bibnamefont {Sung}},\ and\
  \bibinfo {author} {\bibfnamefont {D.~G.}\ \bibnamefont {Whyte}},\ }\bibfield
  {title} {\bibinfo {title} {{ARC: A compact, high-field, fusion nuclear
  science facility and demonstration power plant with demountable magnets}},\
  }\href@noop {} {\bibfield  {journal} {\bibinfo  {journal} {Fusion Engineering
  and Design}\ }\textbf {\bibinfo {volume} {100}},\ \bibinfo {pages} {378}
  (\bibinfo {year} {2015})}\BibitemShut {NoStop}%
\end{thebibliography}%

\appendix

\section{Appendix: Derivation of LCON} \label{app:LCON_derivation}

We derive LCON by setting the net present value (NPV) of a fusion transmutation plant to zero. A plant with capital cost $I^\mathrm{cap}\, P_\mathrm{fus}/10^9$ (where $I^\mathrm{cap}$ is in \$/GW), operating lifetime $L$, availability factor $\mathcal{A}$, and annual discount rate $r$ produces $\dot{N}_\mathrm{n} = P_\mathrm{fus}/E_\mathrm{fus}$ neutrons per second when operating. The capital cost is paid at $t=0$. Annual revenues and operating costs are incurred uniformly over $0 \le t \le L$, with the plant operating a fraction $\mathcal{A}$ of each year.

Each neutron generates revenue $v_\mathrm{n}$ and incurs an operating cost $c_\mathrm{n}^\mathrm{op}$, both in \$/neutron. The annual net cash flow in year $t$ is
\begin{equation}
    F_t = (v_\mathrm{n} - c_\mathrm{n}^\mathrm{op})\, \dot{N}_\mathrm{n}\, T_\mathrm{year}\, \mathcal{A}.
\end{equation}
Because $F_t$ is the same every year (steady-state operation), the NPV is
\begin{equation}
    \mathrm{NPV} = -\frac{I^\mathrm{cap}\, P_\mathrm{fus}}{10^9} + (v_\mathrm{n} - c_\mathrm{n}^\mathrm{op})\, \dot{N}_\mathrm{n}\, T_\mathrm{year}\, \mathcal{A}\, S_\mathrm{disc},
    \label{eq:NPV}
\end{equation}
where $S_\mathrm{disc} = \sum_{t=0}^{L}(1+r)^{-t}$ is the present-value discount sum. Setting $\mathrm{NPV} = 0$ and solving for $v_\mathrm{n}$ gives
\begin{equation}
    v_\mathrm{n}\big|_{\mathrm{NPV}=0} = c_\mathrm{n}^\mathrm{op} + \frac{I^\mathrm{cap}\, E_\mathrm{fus}}{10^9\, T_\mathrm{year}\, \mathcal{A}\, S_\mathrm{disc}} \equiv c_\mathrm{n}^\mathrm{op} + c_\mathrm{n}^\mathrm{cap}.
\end{equation}
This breakeven neutron value is the LCON (\cref{eq:LCON}). The availability $\mathcal{A}$ enters $c_\mathrm{n}^\mathrm{cap}$ because lower availability reduces the total neutron output over which the upfront capital must be amortized. The operating cost $c_\mathrm{n}^\mathrm{op}$ carries no availability or discount factor because it is paid concurrently with neutron production: both the numerator (discounted operating costs) and denominator (discounted neutron output) contain the same $\mathcal{A}\, S_\mathrm{disc}$ factor, which cancels in the ratio. By contrast, the capital cost is paid entirely at $t=0$ and must be amortized over the discounted future neutron output, introducing $\mathcal{A}\, S_\mathrm{disc}$ into $c_\mathrm{n}^\mathrm{cap}$.

The operating cost per neutron is the net electricity cost: the plant purchases electricity at price $\tilde{C}_e$ (\$/J) to power the heating system and receives a credit for electricity generated from blanket heat. The heating power delivered to the plasma is $P_\mathrm{abs} = P_\mathrm{fus}/Q$, requiring a wallplug draw of $P_\mathrm{abs}/(\eta_\mathrm{abs}\,\eta_\mathrm{heat}\,f_h)$. The blanket produces thermal power $\mathcal{K}^* E_\mathrm{fus}\,\dot{N}_\mathrm{n}$, where $\mathcal{K}^* = \mathcal{K} + 1/(\eta_\mathrm{heat}\,Q)$ includes both fusion neutron heating ($\mathcal{K}$) and waste heat from the heating system. Converted to electricity at efficiency $\eta$, the net cost per neutron is
\begin{equation}
    c_\mathrm{n}^\mathrm{op} - v_n^\mathrm{elec}  = \left[\frac{1}{\eta_\mathrm{abs}\,\eta_\mathrm{heat}\,f_h\,Q} - \eta\,\mathcal{K}^*\right] E_\mathrm{fus}\,\tilde{C}_e,
\end{equation}
recovering \cref{eq:vn_elec,eq:cn_op}. At low $Q$ the recirculating power term dominates and $c_\mathrm{n}^\mathrm{op} > 0$ (net electricity consumer); at high $Q$ the blanket electricity credit dominates and $c_\mathrm{n}^\mathrm{op} < 0$ (net electricity producer).

\end{document}